# THE NECESSARY AND SUFFICIENT CONDITIONS FOR TRANSFORMATION FROM DIRAC REPRESENTATION TO FOLDY-WOUTHUYSEN REPRESENTATION

V.P.Neznamov

RFNC-VNIIEF, 607190 Sarov, Nizhniy Novgorod region
e-mail: Neznamov@vniief.ru

The paper describes conditions for transformation from the Dirac representation to the Foldy-Wouthuysen representation. The necessary condition is the block-diagonal transformation of Hamiltonian relative to the upper and lower components of the wave function. The sufficient condition is the wave function transformation law described by relation (6).

It has been demonstrated that the unitary transformations offered by the authors of the papers [14], [15], [16], [17] do not satisfy the sufficiency condition (6) and, hence, they are not the Foldy-Wouthuysen transformations.

In applications, the matrix elements of any operator in the FW representation can be calculated, according to (6), using the normalized two-component wave functions in the Dirac representation known for the given problem.

## 1. Introduction

In the present work, the basic properties of wave functions in the FW representation are determined and the relation between Dirac and FW wave functions is investigated.

The paper determines conditions for transformation of the Dirac representation to the FW representation. The effect of the satisfied conditions is illustrated with many examples. Application of the obtained result allows calculating the expectation values of operators corresponding to the basic classical quantities, because the explicit forms of these operators can be established in the FW representation, but not in the Dirac one.

## 2. The Foldy-Wouthuysen representation

The Foldy-Wouthuysen representation was introduced in the paper [1]. The Foldy-Wouthuysen equation of free motion of a quantum-mechanical particle with spin $\frac{1}{2}$ has the form

$$p_0 \psi_{FW}(x) = (H_0)_{FW} \psi_{FW}(x) = \beta E \psi_{FW}(x). \tag{1}$$

In equation (1) and below the system of units with $\hbar = c = 1$ is used; the inner product of 4-vectors is

$$xy = x^\mu y_\mu = x^0 y^0 - x^k y^k, \mu = 0, 1, 2, 3, k = 1, 2, 3; \ p^\mu = i\frac{\partial}{\partial x_\mu};$$

$\psi_{FW}(x)$ is the four-component wave function;

$\alpha^i = \begin{pmatrix} 0 & \sigma^i \\ \sigma^i & 0 \end{pmatrix}, \beta = \begin{pmatrix} I & 0 \\ 0 & -I \end{pmatrix}$ are Dirac matrices; $\sigma^i$ are Pauli matrices; the operator is

$$E = \sqrt{p^2 + m^2}.$$

The equation (1) solutions are plane waves of both positive and negative energy:

$$\psi_{FW}^{(+)}(x,s) = \frac{1}{(2\pi)^{3/2}} U_s e^{-ipx}, \ \psi_{FW}^{(-)}(x,s) = \frac{1}{(2\pi)^{3/2}} V_s e^{ipx}, \ p_0 = (p^2 + m^2)^{1/2} \tag{2}$$

$U_s = \begin{pmatrix} \varphi_s \\ 0 \end{pmatrix}, V_s = \begin{pmatrix} 0 \\ \chi_s \end{pmatrix}$, $\varphi_s$ and $\chi_s$ in expressions (2) are the two-component Pauli's normalized spin functions.

The following orthonormality and completeness relations are valid for $U_s$ and $V_s$:

$$U_s^\dagger U_{s'} = V_s^\dagger V_{s'} = \delta_{ss'}; U_s^\dagger V_{s'} = V_s^\dagger U_{s'} = 0$$

$$\sum_s (U_s)_\gamma (U_s^\dagger)_\delta = \frac{1}{2}(1+\beta)_{\gamma\delta}; \sum_s (V_s)_\gamma (V_s^\dagger)_\delta = \frac{1}{2}(1-\beta)_{\gamma\delta} \tag{3}$$

In expressions (2) and (3), $\gamma, \delta$ belong to spinor indexes and $s$ belongs to spin indexes. In the further summation with respect to spinor indexes the summation symbol and indexes themselves are not used.

The unitary transformation

$$(H_0)_{FW} = U_0 (H_0)_D U_0^\dagger; \quad U_0 = \sqrt{\frac{E+m}{2E}} \left(1 + \frac{\beta \vec{\alpha} \vec{p}}{E+m}\right) \quad (4)$$

relates Hamiltonian $(H_0)_{FW}$ and free Dirac Hamiltonian $(H_0)_D = \vec{\alpha}\vec{p} + \beta m$ to each other.

The problem of transformation from the Dirac representation to the *FW* representation becomes considerably more complex in a general case of interaction between a fermion and an arbitrary boson field. Eriksen [12] found the exact *FW* transformation for the time-independent Dirac Hamiltonians. Foldy and Wouthuysen [1] found Hamiltonian $H_{FW}$ for interaction with electromagnetic field $A^\mu(\vec{x},t)$ in the form of a series in terms of powers of $\frac{1}{m}$. Blount [2] found Hamiltonian $H_{FW}$ in the form of a series in terms of powers of electromagnetic fields and their time and space derivatives. Case [3] obtained the exact form of the transformation in the presence of the time-independent external magnetic field $\vec{B} = rot\vec{A}$. The Dirac equation is therefore transformed to obtain the following:

$$p_0 \psi_{FW}(x) = H_{FW} \psi_{FW}(x) = \beta \left((\vec{p} - e\vec{A})^2 - e\vec{\sigma}\vec{B} + m^2\right)^{1/2} \psi_{FW}(x). \quad (5)$$

The paper [4] (see also the review [5]) gives the relativistic Hamiltonian in the form of a series in terms of powers of charge *e* for a general case of interaction with an arbitrary external electromagnetic field.

In each of the above-mentioned cases equations for the wave functions in the *FW* representation are non-covariant and their Hamiltonians are non-local (they contain an infinite set of differential operators) and block-diagonal (i.e. they are diagonal relative to the upper and lower components of the wave functions).

The Foldy-Wouthuysen representation, in spite of its non-locality and relative complexity for Hamiltonian $H_{FW}$, is of interest for researchers until now. A number of paradoxes typical for the Dirac representation can be resolved in the *FW* representation. In a free case, the velocity operator in the *FW* representation has a usual form $\vec{v} = \frac{\vec{p}}{E}$ similar to the classical form (in the Dirac representation $\vec{v} = \vec{\alpha}$), electron «trembling» (Zitterbewegung) is absent, the spin operator is kept unchanged with time [1]. The major operators have a considerably simpler form in the *FW* representation. For example, operator $\vec{r}$ in the *FW* representation is a coordinate operator rather than a complex Newton-Wigner operator in the Dirac representation [6]. Similarly, the polarization operator in the *FW* representation is the simple operator $\vec{\Pi} = \beta\vec{\Sigma}$, where $\Sigma^i = \begin{pmatrix} \sigma^i & 0 \\ 0 & \sigma^i \end{pmatrix}$ [7]. The *FW* representation is very useful, if one needs to obtain semi-classical equations for particle motion and spin [8]. The *FW* representation allows easily interpreting the terms of the transformed Dirac Hamiltonian with external fields in the expansions in series in terms of powers of the coupling constant and relativistic parameter $\frac{v}{c}$. Once the relativistic Hamiltonian $H_{FW}$ has been obtained in [4], it becomes possible to consider the quantum-field processes in the *FW* representation [5] within the perturbation theory.

### 3. The necessary and sufficient conditions for transformation from the Dirac representation to the Foldy-Wouthuysen representation

Hamiltonian diagonalization relative to the upper and lower components of the wave function $\psi_D(x)$ is the necessary condition for transformation from the Dirac representation to the Foldy-Wouthuysen representation. However, satisfaction of this condition is not sufficient. As it has been rightly noticed in [9, 10], not each unitary transformation used for diagonalization of the Dirac Hamiltonian leads to the FW representation. It is shown, for example, in [10] that the well-known Eriksen-Kolsrud transformation [11] is not the FW representation.

The second obligatory condition for transition from the Dirac representation to the FW representation is the wave function transformation $\psi_D(\vec{x},t) = \begin{pmatrix} \varphi(\vec{x},t) \\ \chi(\vec{x},t) \end{pmatrix}$ with vanishing the upper and lower components of bi-spinor $\psi_D(x)$ and transforming the normalization operator of the wave function $\psi_D(x)$ into the unit operator. This is a sufficient condition. For a case, when the Dirac Hamiltonian is independent of time (the case of static external fields) this condition can be represented in the following form

$$\psi_D^{(+)}(\vec{x},t) = e^{-iEt} A_+ \begin{pmatrix} \varphi_s^{(+)}(\vec{x}) \\ \chi_s^{(+)}(\vec{x}) \end{pmatrix} \to \psi_{FW}^{(+)}(\vec{x},t) = e^{-iEt} \begin{pmatrix} \varphi_s^{(+)}(\vec{x}) \\ 0 \end{pmatrix}$$

$$\psi_D^{(-)}(\vec{x},t) = e^{iEt} A_- \begin{pmatrix} \varphi_s^{(-)}(\vec{x}) \\ \chi_s^{(-)}(\vec{x}) \end{pmatrix} \to \psi_{FW}^{(-)}(\vec{x},t) = e^{iEt} \begin{pmatrix} 0 \\ \chi_s^{(-)}(\vec{x}) \end{pmatrix}$$

(6)

Here, $E$ is module of the particle energy operator; $A_+$ and $A_-$ are normalization operators, which are not the same, in general, for positive-frequency and negative-frequency solutions; functions $\varphi_s^{(+)}(\vec{x}), \chi_s^{(-)}(\vec{x})$ are normalized to 1. For a free case, $E = (p^2 + m^2)^{1/2}$, $A_+ = A_- = \sqrt{\dfrac{E+m}{2E}}$; for the positive-energy solutions we have $\varphi_s^{(+)}(\vec{x}) = e^{i\vec{p}\vec{x}} \cdot \varphi_s$ and for the negative-energy solutions - $\chi_s^{(-)}(\vec{x}) = e^{-i\vec{p}\vec{x}} \cdot \chi_s$; $\varphi_s$ and $\chi_s$ are the two-component Pauli's spin functions (see (2)).

Functions $\varphi_s(\vec{x})$ and $\chi_s(\vec{x})$ for a particle moving in static external fields are the appropriate eigen-functions of the Dirac equation. The sufficiency condition implies transformation of the wave functions to the form (6) with the normalization operator equal to 1.

In general, if the Dirac and FW Hamiltonians depend on time, the sufficiency condition (6) has the same meaning, because we use expansions in the Dirac equation solutions obtained either for free motion of particles, or for motion in the presence of static external boson fields when solving the particular problems of physics (at least, with the use of the perturbation theory).

The sufficiency criterion validity can be proved using the Eriksen transformation [12], which is the exact FW transformation for the time-independent Hamiltonians $H_D$. The transformation matrix $U_{Er}$ may be represented, in general, as a product of two factors

$$U_{Er} \equiv U_{FW} = \frac{1}{2}(1+\beta\lambda)\left(\frac{1}{2}+\frac{\beta\lambda+\lambda\beta}{4}\right)^{-\frac{1}{2}}; \qquad (7)$$

$$\lambda = \frac{H_D}{\left(H_D^{\,2}\right)^{\frac{1}{2}}}, \quad \lambda = \pm 1 \quad \text{for the positive-energy and negative-energy solutions,}$$

respectively,

where $H_D$ is Hamiltonian in the Dirac representation. It is important that [12]

$$\lambda^2 = 1, [\beta\lambda, \lambda\beta] = 0 \qquad (8)$$

and the operator $\beta\lambda + \lambda\beta$ is even,

$$[\beta,(\beta\lambda+\lambda\beta)] = 0. \qquad (9)$$

Consider the normalized Dirac wave function, for example, for the positive-energy solutions.

$$\psi_D^{(+)}(x) = e^{-iEt} A_+ \begin{pmatrix} \varphi^{(+)}(\vec{x}) \\ \chi^{(+)}(\vec{x}) \end{pmatrix} \qquad (10)$$

Definition of the normalized operator $A_+$ in (10) implies that spinor $\varphi^{(+)}(\vec{x})$ is normalized to 1 $\left(\int \varphi^{(+)\dagger}(\vec{x})\varphi^{(+)}(\vec{x})dV = 1\right)$.

Apply the Eriksen transformation to equation (10) and obtain

$$\psi_{FW}^{(+)}(x) = U_{Er} \cdot \psi_D = \left[\frac{1}{2}+\frac{1}{4}(\beta\lambda+\lambda\beta)\right]^{-\frac{1}{2}} \frac{1}{2}(1+\beta\lambda)\psi_D^{(+)}(x). \qquad (11)$$

The Eriksen transformation in equation (11), in contrast to equation (7), has the changed order of factors, which is admissible in view of relations (8).

Since $\lambda\psi_D^{(+)}(x) = 1$, expression (11) takes the form

$$\psi_{FW}^{(+)}(x) = e^{-iEt}\left[\frac{1}{2}+\frac{1}{4}(\beta\lambda+\lambda\beta)\right]^{-\frac{1}{2}} A_+ \begin{pmatrix} \varphi^{(+)}(\vec{x}) \\ 0 \end{pmatrix}. \qquad (12)$$

The wave function normalization requirement may be written as

$$\int \psi_{FW}^{(+)\dagger}(x)\psi_{FW}^{(+)}(x)dV = \int \varphi^{(+)\dagger}(\vec{x})A_+\left(\frac{1}{2}+\frac{1}{4}(\beta\lambda+\lambda\beta)\right)^{-1} A_+\varphi^{(+)}(\vec{x})dV = 1 \ . \qquad (13)$$

Meeting the requirement (13) means necessarily meeting the requirement

$$A_+\left(\frac{1}{2}+\frac{1}{4}(\beta\lambda+\lambda\beta)\right)^{-1} A_+ = 1. \qquad (14)$$

Multiply the left-hand and right-hand sides of equation (14) by the operator $A_+^{-1}$ and obtain

$$\left(\frac{1}{2}+\frac{1}{4}(\beta\lambda+\lambda\beta)\right)^{-1}=A_+^{-2};$$
$$\left(\frac{1}{2}+\frac{1}{4}(\beta\lambda+\lambda\beta)\right)^{-\frac{1}{2}}A_+=1 \qquad (15)$$

Expressions (15) prove the sufficiency condition (6).

According to (12) we, indeed, obtain

$$\psi_{FW}^{(+)}(x)=e^{-iEt}\begin{pmatrix}\varphi^{(+)}(x)\\ 0\end{pmatrix} \qquad (16)$$

Similar considerations are valid for the negative-energy solutions and the result is

$$\psi_{FW}^{(-)}(x)=e^{iEt}\begin{pmatrix}0\\ \chi^{(-)}(\vec{x})\end{pmatrix} \qquad (17)$$

In the next section, the sufficiency condition (6) is applied to some block-diagonal Hamiltonians known to the given paper author and obtained by unitary transformations of the Dirac representation. It is shown that the sufficiency condition (6) is, sometimes, not satisfied and, if it is the case, the unitary transformations performed are not the Foldy-Wouthuysen transformations.

**4. Consideration of the sufficiency condition (6) for some unitarily transformed wave functions.**

**4.1. Free motion**

$$H_D = \vec{\alpha}\vec{p} + \beta m; U_{FW} = \sqrt{\frac{E+m}{2E}}\left(1 + \frac{\beta\vec{\alpha}\vec{p}}{E+m}\right); H_{FW} = \beta E; E = \sqrt{\vec{p}^2 + m^2};$$

$$\psi_D^{(+)}(\vec{x},t) = \sqrt{\frac{E+m}{2E}} e^{-iEt} \begin{pmatrix} \varphi_s^{(+)}(\vec{x}) \\ \frac{\vec{\sigma}\vec{p}}{E+m}\varphi_s^{(+)}(\vec{x}) \end{pmatrix}; \psi_{FW}^{(+)}(\vec{x},t) = U_{FW}\psi_D^{(+)}(\vec{x},t) = e^{-iEt}\begin{pmatrix} \varphi_s^{(+)}(\vec{x}) \\ 0 \end{pmatrix}; \quad (18)$$

$$\psi_D^{(-)}(\vec{x},t) = \sqrt{\frac{E+m}{2E}} e^{iEt} \begin{pmatrix} -\frac{\vec{\sigma}\vec{p}}{E+m}\chi_s^{(-)}(\vec{x}) \\ \chi_s^{(-)}(\vec{x}) \end{pmatrix}; \psi_{FW}^{(-)}(\vec{x},t) = U_{FW}\psi_D^{(-)}(\vec{x},t) = e^{iEt}\begin{pmatrix} 0 \\ \chi_s^{(-)}(\vec{x}) \end{pmatrix}.$$

### 4.2. Motion in static magnetic field

$$H_D = \vec{\alpha}\vec{\pi} + \beta m; U_{FW} = \sqrt{\frac{E+m}{2E}}\left(1 + \frac{\beta\vec{\alpha}\vec{\pi}}{E+m}\right); H_{FW} = \beta E; \vec{\pi} = \vec{p} - e\vec{A}(\vec{x});$$

$$E = \sqrt{(\vec{p} - e\vec{A})^2 - e\vec{\sigma}\vec{B} + m^2};$$

$$\psi_D^{(+)}(\vec{x},t) = \sqrt{\frac{E+m}{2E}} e^{-iEt} \begin{pmatrix} \varphi_s^{(+)}(\vec{x}) \\ \frac{\vec{\sigma}\vec{\pi}}{E+m}\varphi_s^{(+)}(\vec{x}) \end{pmatrix}; \psi_{FW}^{(+)}(\vec{x},t) = U_{FW}\psi_D^{(+)}(\vec{x},t) = e^{-iEt}\begin{pmatrix} \varphi_s^{(+)}(\vec{x}) \\ 0 \end{pmatrix}; \quad (19)$$

$$\psi_D^{(-)}(\vec{x},t) = \sqrt{\frac{E+m}{2E}} e^{iEt} \begin{pmatrix} -\frac{\vec{\sigma}\vec{\pi}}{E+m}\chi_s^{(-)}(\vec{x}) \\ \chi_s^{(-)}(\vec{x}) \end{pmatrix}; \psi_{FW}^{(-)}(\vec{x},t) = U_{FW}\psi_D^{(-)}(\vec{x},t) = e^{iEt}\begin{pmatrix} 0 \\ \chi_s^{(-)}(\vec{x}) \end{pmatrix}$$

### 4.3. Motion in static electrical field

Here we demonstrate, how the condition (6) is satisfied, by using only terms ~ $e$ and terms including the quadratic ones with respect to $\frac{v}{c}$ in the $U_{FW}$ expansion in series in terms of powers of the charge $e$ [4]. Apparently, the procedure can be implemented for any order of expansion in terms of $e$ and $\frac{v}{c}$, if we use the mathematical apparatus from [4].

With denotations used in [4], we obtain, within the accepted accuracy, that

$$H_D = \vec{\alpha}\vec{p} + \beta m + eA_0(\vec{x}); \quad U_{FW} = \left(1+\delta_1^0 + \delta_1^e\right)U_0 = 1 + \frac{\beta\vec{\alpha}\vec{p}}{2m} - \frac{1}{8}\frac{p^2}{m^2} - \frac{ie}{4m^2}(\vec{\alpha}\nabla A_0) -$$

$$-\frac{i\beta}{16m^3}\left((\vec{\alpha}\vec{p})(\vec{\alpha}\nabla A_0) - (\vec{\alpha}\nabla A_0)(\vec{\alpha}\vec{p})\right); \quad H_{FW} = \beta E;$$

$$E = m + \frac{p^2}{2m} + e\beta\left(A_0 + \frac{i}{8m^2}\left((\vec{\alpha}\vec{p})(\vec{\alpha}\nabla A_0) - (\vec{\alpha}\nabla A_0)(\vec{\alpha}\vec{p})\right)\right);$$

$$\psi_D^{(+)}(\vec{x},t) = e^{-iEt}\left(1 - \frac{1}{8}\frac{p^2}{m^2} - \frac{ie}{16m^3}\left(\vec{\sigma}\vec{p}(\vec{\sigma}\nabla A_0) - (\vec{\sigma}\nabla A_0)(\vec{\sigma}\vec{p})\right)\right)\begin{pmatrix} \varphi_s^{(+)}(\vec{x}) \\ \left(\frac{\vec{\sigma}\vec{p}}{2m} + \frac{ie\vec{\sigma}\nabla A_0}{4m^2}\right)\varphi_s^{(+)}(\vec{x}) \end{pmatrix}$$

$$\psi_{FW}^{(+)}(\vec{x},t) = U_{FW}\psi_D^{(+)}(\vec{x},t) = e^{-iEt}\begin{pmatrix} \varphi_s^{(+)}(\vec{x}) \\ 0 \end{pmatrix}$$

$$\psi_D^{(-)}(\vec{x},t) = e^{iEt}\left(1 - \frac{1}{8}\frac{p^2}{m^2} + \frac{ie}{16m^3}\left(\vec{\sigma}\vec{p}\vec{\sigma}\nabla A_0 - \vec{\sigma}\nabla A_0\vec{\sigma}\vec{p}\right)\right)\begin{pmatrix} -\left(\frac{\vec{\sigma}\vec{p}}{2m} - \frac{ie\vec{\sigma}\nabla A_0}{4m^2}\right)\chi_s^{(-)}(\vec{x}) \\ \chi_s^{(-)}(\vec{x}) \end{pmatrix}$$

$$\psi_{FW}^{(-)}(\vec{x},t) = U_{FW}\psi_D^{(-)}(\vec{x},t) = e^{iEt}\begin{pmatrix} 0 \\ \chi_s^{(-)}(\vec{x}) \end{pmatrix} \quad (20)$$

In expressions (18)-(20), $\vec{p}$ and functions $f(\vec{p})$ imply the corresponding operators; $\varphi_s(\vec{x})$, $\chi_s(\vec{x})$ are the two-component functions satisfying the equation

$$\left[p_0 - eA_0 - \beta m - \vec{\sigma}\left(\vec{p} - e\vec{A}\right)(p_0 - eA_0 + \beta m)^{-1}\vec{\sigma}\left(\vec{p} - e\vec{A}\right)\right]\psi_D = 0 \quad (21)$$

In (21) we have $\psi = \begin{pmatrix}\varphi_s \\ \chi_s\end{pmatrix}$; equations for $\varphi_s$ and $\chi_s$ are obtained with $\beta = I$ and $\beta = -I$, respectively, (see [9, 4]); for static external fields we have $p_0\psi = \varepsilon\psi$; $\left(\psi(\vec{x},t) = e^{-i\varepsilon t}\begin{pmatrix}\varphi_s(\vec{x}) \\ \chi_s(\vec{x})\end{pmatrix}\right)$; for $\beta = +I$ $\varepsilon = E$; for $\beta = -I$ $\varepsilon = -E$.

For convenience, the particle energy operators of different forms in expressions (18)-(20) are denoted with the same symbol $E$.

In (20) we suppose that in operator $E$ $\beta = +1$ for the positive-frequency solutions $\psi^{(+)}(\vec{x},t)$ and $\beta = -1$ for the negative-frequency solutions $\psi^{(-)}(\vec{x},t)$ (see [9, 4]).

It follows from (18)-(20) that the sufficiency condition (6) is satisfied in each of the cases above and, thus, the given unitary transformations are the Foldy-Wouthuysen transformations.

### 4.4. Super-algebra in the Dirac equation with static external fields

Using the super-symmetric quantum mechanics concepts, the paper [13] discusses a wide range of interactions between a Dirac particle and static external fields, which provides the closed form of the Dirac Hamiltonian diagonalization. The paper [13] authors use $SU(2)$ transformation of the Dirac Hamiltonian, as the Foldy-Wouthuysen representation.

It is interesting to know, how the sufficiency condition (6) is satisfied for this transformation. Using denotations from [13], we have $H_D = Q + Q^\dagger + \lambda$, where $\lambda$ is the Hermitian operator, $Q$ and $Q^\dagger$ are the two fermion operators meeting the following requirements: $Q^2 = Q^{\dagger 2} = 0, \{Q,\lambda\} = \{Q^\dagger,\lambda\} = 0$.

Then, the Hermitian operators of $SU(2)$-algebra are introduced:

$$J_1 = \frac{Q+Q^\dagger}{2(\{Q,Q^\dagger\})^{1/2}}; J_2 = \frac{-i\lambda(Q+Q^\dagger)}{2(\lambda^2\{Q,Q^\dagger\})^{1/2}}; J_3 = \frac{\lambda}{2(\lambda^2)^{1/2}}; [J_i, J_j] = i\varepsilon_{ijk}J_k$$

The transformation operator is

$$U_{FW} = e^{iJ_2\theta} = \cos\frac{\theta}{2} + 2iJ_2\sin\frac{\theta}{2} = \sqrt{\frac{1}{2}(1+\cos\theta)} + \frac{\lambda(Q+Q^\dagger)}{(\lambda^2\{Q,Q^\dagger\})^{1/2}}\sqrt{\frac{1}{2}(1-\cos\theta)} \qquad (22)$$

In contrast to [13], exponent $iJ_2\theta$ in (22) is taken with its positive sign. As one can see below, this is a necessary step for the sufficiency condition (6) to be satisfied.

$$H_{FW} = e^{iJ_2\theta}H_D e^{-iJ_2\theta} = (Q+Q^\dagger)\cos\theta + 2iJ_2\lambda\sin\theta + \lambda\cos\theta + 2iJ_2(Q+Q^\dagger)\sin\theta \qquad (23)$$

If $tg\theta = \frac{\{Q,Q^\dagger\}^{1/2}}{(\lambda^2)^{1/2}}$ $\left(\sin\theta = \frac{(\{Q,Q^\dagger\})^{1/2}}{(\{Q,Q^\dagger\}+\lambda^2)^{1/2}}, \cos\theta = \frac{(\lambda^2)^{1/2}}{(\{Q,Q^\dagger\}^{1/2}+\lambda^2)^{1/2}}\right)$,

expression (23) is reduced to its diagonal form

$$H_{FW} = \frac{\lambda}{(\lambda^2)^{1/2}}(\{Q,Q^\dagger\}+\lambda^2)^{1/2}. \qquad (24)$$

If $\lambda = \beta m$; $Q = \begin{pmatrix} 0 & 0 \\ M & 0 \end{pmatrix}$, $Q^\dagger = \begin{pmatrix} 0 & M^\dagger \\ 0 & 0 \end{pmatrix}$;

$$M = \vec{\sigma}(\vec{p} + \vec{C}) - iC_5; \quad C_I = A_I - iE_I, \quad I = 1,2,3,5,$$

the Dirac Hamiltonian $H_D$ can be written as

$$H_D = \vec{\alpha}\vec{\pi} + i\beta\gamma_5\pi_5 + \beta m; \quad \pi_I = p_I + A_I(\vec{x}) + i\beta E_I(\vec{x}), \quad I = 1,2,3,5, \quad p_5 = 0. \quad (25)$$

The following interactions are presented using these denotations:

$A_5$ is the pseudo-scalar potential, $E_5$ is the time component of the axial-vector potential, $\vec{E}$ is the "electrical" component of interaction of the abnormal magnetic moment of a particle, $\vec{A}$ is the minimum magnetic interaction; if $A_5 = E_5 = 0, \vec{A} = 0, \vec{E} = \vec{r}$, Hamiltonian $H_D$ is reduced to Hamiltonian of the Dirac oscillator. Each of the interactions above admits closed transformation to the diagonalized form (24).

Let us check, whether the $SU(2)$-transformation (22) is the FW transformation with the satisfied sufficiency condition (6).

So, $H_D = Q + Q^\dagger + \beta m$, where $Q$ and $Q^\dagger$ are defined in (25).

$$H_{FW} = \beta E = \beta\left(\{Q,Q^\dagger\} + m^2\right)^{1/2}; \quad E^2 = \{Q,Q^\dagger\} + m^2 = \begin{pmatrix} M^\dagger M + m^2 & 0 \\ 0 & MM^\dagger + m^2 \end{pmatrix}$$

$$U_{FW} = \sqrt{\frac{E+m}{2E}}\left(1 + \frac{\beta(Q+Q^\dagger)}{E+m}\right);$$

$$\psi_D^{(+)}(\vec{x},t) = e^{-iEt}\sqrt{\frac{E+m}{2E}}\begin{pmatrix} \varphi_s^{(+)}(\vec{x}) \\ \frac{1}{E+m}M\varphi_s^{(+)}(\vec{x}) \end{pmatrix}; \quad \psi_{FW}^{(+)}(\vec{x},t) = U_{FW}\psi_D^{(+)}(\vec{x},t) = e^{-iEt}\begin{pmatrix} \varphi_s^{(+)}(\vec{x}) \\ 0 \end{pmatrix} \quad (26)$$

$$\psi_D^{(-)}(\vec{x},t) = e^{iEt}\sqrt{\frac{E+m}{2E}}\begin{pmatrix} -\frac{1}{E+m}M^\dagger\chi_s^{(-)}(\vec{x}) \\ \chi_s^{(-)}(\vec{x}) \end{pmatrix}; \quad \psi_{FW}^{(-)}(\vec{x},t) = U_{FW}\psi_D^{(-)}(\vec{x},t) = e^{iEt}\begin{pmatrix} 0 \\ \chi_s^{(-)}(\vec{x}) \end{pmatrix}$$

One can see from expressions (26) that, indeed, the $SU(2)$- transformation in [13] (with the changed sign of exponent $iJ_2\theta$) provides satisfaction of the sufficiency condition (6) and is the FW transformation at the same time. If the authors' sign of the exponent in expression (22) remains unchanged $\left(e^{-iJ_2\theta}, see [13]\right)$, the transformation in (26) becomes equal to $U = \sqrt{\frac{E+m}{2E}}\left(1 - \frac{\beta(Q+Q^\dagger)}{E+m}\right)$. Such transformation does not provide satisfaction of the sufficiency condition (6), though the requirement of block-diagonalization of Hamiltonian can be met (see [13]).

## 4.5. A Dirac particle's motion in external gravitational field (Obukhov Hamiltonian [14, 15]).

The papers [14, 15] offer the Dirac Hamiltonian of the form

$$H_D = \beta m V + \frac{1}{2}\left\{(\vec{\alpha}\vec{p}), \frac{V}{W}\right\} \quad (27)$$

for static metrics $dS^2 = V^2(\vec{x})(dx^0)^2 - W^2(\vec{x})d\vec{x}^2$. Then, diagonalization of the Hamiltonian (27) using the closed Eriksen-Kolsrud transformation is performed in [14, 15]. The paper [10] proves nonequivalence of the FW and Eriksen-Kolsrud transformations and offers the explicit form of the operator of transition between the FW and Eriksen-Kolsrud representations for the free Dirac equation.

Let us, first, check, whether the sufficiency condition (6) is satisfied for the Obukhov Hamiltonian (27), using the method described in [4].

For a free case, $H_D = \vec{\alpha}\vec{p} + \beta m; H_{E-K} = \beta E;$ the transformation matrix used in [14] has the form

$$U_{E-K} = \frac{1}{\sqrt{2}}(1+\beta i\gamma_5\beta)\frac{1}{\sqrt{2}}\left(1+i\gamma_5\beta\frac{\vec{\alpha}\vec{p}+\beta m}{E}\right) = \frac{1}{2}\left(1+\frac{m}{E}+\frac{i\beta\vec{\sigma}\vec{p}}{E}-i\gamma_5+i\gamma_5\frac{m}{E}-\frac{\beta\gamma_5\vec{\sigma}\vec{p}}{E}\right);$$

$$\gamma_5 = i\alpha_1\alpha_2\alpha_3;$$

$$\psi_{E-k}^{(+)}(\vec{x},t) = U_{E-k} \cdot \psi_D^+(\vec{x},t) = e^{-iEt}\sqrt{\frac{E+m}{2E}}\begin{pmatrix}\left(1+\frac{i\vec{\sigma}\vec{p}}{E+m}\right)\varphi_s^{(+)}(\vec{x})\\0\end{pmatrix}; \quad (28)$$

$$\psi_{E-k}^{(-)}(\vec{x},t) = U_{E-k} \cdot \psi_D^{(-)}(\vec{x},t) = e^{iEt}\sqrt{\frac{E+m}{2E}}\begin{pmatrix}0\\\left(1-\frac{i\vec{\sigma}\vec{p}}{E+m}\right)\chi_s^{(-)}(\vec{x})\end{pmatrix};$$

It can be seen from (28) that the condition (6) is not satisfied and one needs to perform additional transformation $U_{E-k\to FW}$, which reduces the wave functions to the form (6) with no changes of the form of Hamiltonian $H_{E-k} = H_{FW} = \beta E$.

This transformation looks like [10], as follows:

$$U_{E-k \to FW} = \sqrt{\frac{E+m}{2E}} \left(1 - \frac{i\beta\vec{\sigma}\vec{p}}{E+m}\right). \tag{29}$$

Now, let us try to transform the Hamiltonian (27) with the method described in [4, 5] and check, whether the sufficiency condition is satisfied, or not. Similar to the papers [14, 10], our consideration is performed to the first-order accuracy of potentials $(V-1), (F-1)$ and their first derivatives in space coordinates; in the expansions in series in terms of relativistic parameter $\frac{v}{c}$ we restrict ourselves with values including $\frac{v^2}{c^2}$.

Using denotations from [4, 5] we have

$$U_{FW} = (1 + \delta_1^0 + \delta_1^e) U_0; \quad U_0 = \sqrt{\frac{E+m}{2E}} \left(1 + \frac{\beta\vec{\alpha}\vec{p}}{E+m}\right); \quad E = \sqrt{p^2 + m^2};$$

$$\delta_1^0 \beta E - \beta E \cdot \delta_1^0 + N = 0; \quad \delta_1^e R + R \delta_1^e = RL\delta_1^0 - \delta_1^0 LR; \quad R = \sqrt{\frac{E+m}{2E}};$$

$$L = \frac{\beta\vec{\alpha}\vec{p}}{E+m};$$

$$H_{FW} = \beta E + K_1; \quad K_1 = \delta_1^e \beta E - \beta E \delta_1^e + C;$$

The Hamiltonian (27) is written as

$$H_D = \vec{\alpha}\vec{p} + \beta m + \mathcal{E} + Q; \quad \mathcal{E} = \beta m(V-1); \quad Q = \frac{1}{2}\{(F-1), \vec{\alpha}\vec{p}\}; \quad F = \frac{V}{W}.$$

Hence, within the accepted accuracy [4, 5] we have

$$N = \frac{\beta}{2m}(\vec{\alpha}\vec{p}\mathcal{E} - \mathcal{E}\vec{\alpha}\vec{p}) + Q = \frac{1}{2}\{(V-1), \vec{\alpha}\vec{p}\} + \frac{1}{2}\{(F-1), \vec{\alpha}\vec{p}\}; \quad \delta_1^0 = \frac{\beta}{2}N.$$

$$\delta_1^e = \frac{1}{16m^2}\left((F-V)p^2 - p^2(F-V)\right);$$

$$U_{FW} = (1 + \delta_1^0 + \delta_1^e)\left(1 + \frac{\beta\vec{\alpha}\vec{p}}{2m} - \frac{1}{8}\frac{p^2}{m^2}\right) = 1 + \frac{\beta\vec{\alpha}\vec{p}}{2m} - \frac{1}{8}\frac{p^2}{m^2} + \frac{\beta}{4m}\{(F-V), \vec{\alpha}\vec{p}\} -$$

$$-\frac{1}{16m^2}\left((F-V)p^2 + 2\vec{\alpha}\vec{p}(F-V)\vec{\alpha}\vec{p} + p^2(F-V)\right)$$

$$H_{FW} = \beta E + K_1 = \beta m + \frac{\beta p^2}{2m} + C = \beta m + \frac{\beta p^2}{2m} + \beta m(V-1) - \frac{\beta}{4m}\{p^2 V - 1\} +$$

$$+ \frac{\beta}{2m}\{p^2 F - 1\} + \frac{\beta}{4m}\left[2\vec{\sigma}\left(\vec{f} \times \vec{p}\right) + \nabla \vec{f}\right] - \frac{\beta}{8m}\left[2\vec{\sigma}\left(\vec{\Phi} \times \vec{p}\right) + \nabla \vec{\Phi}\right] = \beta E' \quad (30)$$

In expressions (30), $\vec{\Phi} = \nabla V$; $\vec{f} = \nabla F$.

The expression for $H_{FW}$ is identical to the expression obtained in [10] (see expression (14)).

The Dirac equation with Hamiltonian (27) for static potentials $V(\vec{x}), W(\vec{x})$ can be written as

$$\psi_D(\vec{x},t) = e^{-i\varepsilon t}\begin{pmatrix}\varphi_s(\vec{x})\\ \chi_s(\vec{x})\end{pmatrix}; \quad \begin{cases}\varepsilon\varphi_s = m\varphi_s + m(V-1)\varphi_s + \left(\vec{\sigma}\vec{p} + \frac{1}{2}\{\vec{\sigma}\vec{p},(F-1)\}\right)\chi_s \\ \varepsilon\chi_s = -m\chi_s - m(V-1)\chi_s + \left(\vec{\sigma}\vec{p} + \frac{1}{2}\{\vec{\sigma}\vec{p},(F-1)\}\right)\varphi_s\end{cases} \quad (31)$$

It follows from the second equation of the equation system (31) that

$$\chi_s = \frac{1}{\varepsilon + m + m(V-1)}\left(\vec{\sigma}\vec{p} + \frac{1}{2}\{\vec{\sigma}\vec{p},(F-1)\}\right)\cdot\varphi_s =$$

$$= \left(\vec{\sigma}\vec{p} + \frac{1}{2}\{\vec{\sigma}\vec{p},(F-1)\}\right)\frac{1}{\varepsilon + m + m(V-1)}\varphi_s - [m(V-1),\vec{\sigma}\vec{p}]\frac{1}{(\varepsilon+m)^2}\varphi_s \quad (32)$$

For $\varepsilon > 0$, the replacement $\varepsilon\varphi_s^{(+)} = E'\varphi_s^{(+)}$ can be performed in (32).

Hence, within the accepted accuracy we have

$$\chi_s^{(+)} = \left(\frac{\vec{\sigma}\vec{p}}{2m} + \frac{1}{4m}\{(F-V),\vec{\sigma}\vec{p}\}\right)\varphi_s^{(+)} \quad (33)$$

Similarly, for $\varepsilon < 0$ $\varepsilon\chi_s^{(-)} = -E'\chi_s^{(-)}$ and $\varphi_s^{(-)} = -\left(\frac{\vec{\sigma}\vec{p}}{2m} + \frac{1}{4m}\{(F-V),\vec{\sigma}\vec{p}\}\right)\chi_s^{(-)} \quad (34)$

Using expressions (30)÷(34) and the wave function normalization requirements, we obtain:

$$\psi_D^{(+)}(\vec{x},t) = e^{-iE't}A_+\begin{pmatrix}\varphi_s^{(+)}(\vec{x})\\ \chi_s^{(+)}(\vec{x})\end{pmatrix} =$$

$$= e^{-iE't}\left(1 - \frac{1}{8}\frac{p^2}{m^2} - \frac{1}{16m^2}\left((F-V)p^2 + 2\vec{\sigma}\vec{p}(F-V)\vec{\sigma}\vec{p} + p^2(F-V)\right)\right)\times$$

$$\times\begin{pmatrix}\varphi_s(x)\\ \left(\frac{\vec{\sigma}\vec{p}}{2m} + \frac{1}{4m}\{(F-V),\vec{\sigma}\vec{p}\}\right)\varphi_s^{(+)}(\vec{x})\end{pmatrix};$$

$$\psi_D^{(-)}(\vec{x},t) = e^{iE't} A_- \begin{pmatrix} \varphi_s^{(-)}(\vec{x}) \\ \chi_s^{(-)}(\vec{x}) \end{pmatrix} =$$

$$= e^{iE't} \left(1 - \frac{1}{8}\frac{p^2}{m^2} - \frac{1}{16m^2}\left((F-V)p^2 + 2\vec{\sigma}\vec{p}(F-V)\vec{\sigma}\vec{p} + p^2(F-V)\right)\right) \times$$

$$\times \begin{pmatrix} -\left(\frac{\vec{\sigma}\vec{p}}{2m} + \frac{1}{4m}\{(F-V),\vec{\sigma}\vec{p}\}\right)\chi_s^{(-)}(\vec{x}) \\ \chi_s^{(-)}(\vec{x}) \end{pmatrix};$$

$$\psi_{FW}^{(+)}(\vec{x},t) = U_{FW}\psi_D^{(+)}(\vec{x},t) = e^{-iE't}\begin{pmatrix}\varphi_s^{(+)}(\vec{x}) \\ 0\end{pmatrix};$$

$$\psi_{FW}^{(-)}(\vec{x},t) = U_{FW}\psi_D^{(-)}(\vec{x},t) = e^{iE't}\begin{pmatrix}0 \\ \chi_s^{(-)}(\vec{x})\end{pmatrix}$$

(35)

Expression (35) shows that the sufficiency condition (6) is satisfied.

And vice versa, application of the Obukhov transformation diagonalizing the Hamiltonian (27) to the wave functions $\psi_D^{(+)}(\vec{x},t)$, $\psi_D^{(-)}(\vec{x},t)$ doesn't provide satisfaction of the sufficiency condition (6) (similar to the case of free motion (28)).

Really, using denotations from [14] the Obukhov transformation can be written as

$$U_{E-K} = \frac{1}{\sqrt{2}}(1+\beta J)\frac{1}{\sqrt{2}}(1+J\hat{\Lambda}) = \frac{1}{2}(1-i\gamma_5)\left(1+i\gamma_5\cdot\beta\frac{H_D}{(H_D^2)^{1/2}}\right)\frac{1}{2}(1+i\gamma_5)(1-i\gamma_5) \quad (36)$$

Within the accepted accuracy (see also [14], [10]), expression (36) can be written as

$$U_{E-K} = \frac{1}{2}\left(1+i\gamma_5 - \frac{\beta\gamma_5}{2}\{\vec{\sigma}\vec{p},F\}\frac{1}{mV} - \frac{i\gamma_5}{4m^2}\left(\frac{1}{W}p^2F + F\cdot p^2\frac{1}{W}\right)\frac{1}{V} - \frac{i\gamma_5}{4m^2}\left(\nabla\vec{f} + 2\vec{\sigma}\left(\vec{f}\times\vec{p}\right)\right) - \frac{i\gamma_5\cdot\beta}{2m}\vec{\sigma}\vec{\Phi} - \frac{\gamma_5}{4m^2}\{\vec{\sigma}\vec{p},F\}\vec{\sigma}\vec{\Phi}\right)(1-i\gamma_5)$$

(37)

Hence, according to (35) we have

$$\psi_{E-K}^{(+)}(\vec{x},t) = U_{E-K}\cdot\psi_D^{(+)}(\vec{x},t) =$$

$$= e^{-iE't}\left(1 - \frac{p^2}{8m^2} - \frac{1}{16m^2}\left((F-V)p^2 + 2\vec{\sigma}\vec{p}(F-V)\vec{\sigma}\vec{p} + p^2(F-V)\right) + \frac{i}{4}\{\vec{\sigma}\vec{p},F\}\frac{1}{mV} - \frac{1}{4m}\vec{\sigma}\vec{\Phi}\right)\times$$

$$\times \begin{pmatrix}\varphi_s^{(+)}(\vec{x}) \\ 0\end{pmatrix}$$

(38)

$$\psi_{E-K}^{(-)}(\vec{x},t) = U_{E-K} \cdot \psi_D^{(-)}(\vec{x},t) =$$

$$= e^{iE't}\left(1 - \frac{p^2}{8m^2} - \frac{1}{16m^2}\left((F-V)p^2 + 2\vec{\sigma}\vec{p}(F-V)\vec{\sigma}\vec{p} + p^2(F-V)\right) - \frac{i}{4}\{\vec{\sigma}\vec{p},F\}\frac{1}{mV} - \frac{1}{4m}\vec{\sigma}\vec{\Phi}\right) \times$$

$$\times \begin{pmatrix} 0 \\ \chi_s^{(-)}(\vec{x}) \end{pmatrix}$$

(39)

One can see that the sufficiency condition (6) is not satisfied in expressions (38), (39). For free motion $(F=1, V=1)$, the wave functions in (38), (39) are identical to the wave functions in (28), within the accepted accuracy.

The closed transformation of the Eriksen-Kolsrud type was fulfilled in [16] with the Obukhov Hamiltonian using the super-symmetric quantum mechanics methods, the resultant Hamiltonian coincides with the transformed Hamiltonian from [14] in each order of the expansion in terms of powers of $\frac{1}{m}$.

The transformation matrix with denotations from [16] looks like

$$U = \frac{1}{\sqrt{2}}\left(1 + \beta\frac{Q}{(Q^2)^{1/2}}\right)\frac{1}{\sqrt{2}}(1 - i\gamma_5), \quad Q = \frac{1}{2}\{\vec{\alpha}\vec{p},F\} + i\gamma_5\beta mV \qquad (40)$$

If we expand in series the expression (40) with the first-order accuracy of potentials $(V-1), (F-1)$ and their first space derivatives and restrict ourselves to $\sim \frac{1}{m^2}$ powers in the expansion in terms of mass, we can see that (40) coincides with the Obukhov transformation matrix (37) within the same accuracy.

Hence, the transformation fulfilled by the authors of paper [16] is not the Foldy-Wouthuysen transformation, because the sufficiency condition (6) is not satisfied, similar to the transformation fulfilled in [14].

In [17] the authors also use the Eriksen-Kolsrud transformation, they call it the exact FW transformation and don't check the wave function transformation law. In view of the foresaid, the Eriksen-Kolsrud transformation is not the Foldy-Wouthuysen transformation, because the wave function transformation condition (6) is not satisfied.

## 5. Conclusion

The paper states and proves the necessary and sufficient conditions for transition from the Dirac representation to the Foldy-Wouthuysen representation. The exact connection between the wave functions in the both representations has been found. It has been demonstrated by a number of examples that in some cases ([14], [15], [16], [17]) block-diagonalization of the Dirac Hamiltonian is insufficient for transformation to the Foldy-Wouthuysen representation; such transformation necessarily requires satisfaction of the sufficiency condition (6). The results obtained allow unambiguous transformation to the FW representation and calculation of the matrix elements of operators corresponding to the main classical quantities, because the exact form of these operators can be easily found in the Foldy-Wouthuysen representation, in contrast to the Dirac representation.

In applications, during calculation of the matrix elements of operators in the Foldy-Wouthuysen representation it is sufficient, according to (6), to use the upper or lower components of the normalized Dirac wave functions.

$$\begin{array}{l}\left\langle \psi_{FW}^{(+)\dagger}(\vec{x}) \big| A_{FW} \big| \psi_{FW}^{(+)}(\vec{x}) \right\rangle = \left\langle \varphi_{D}^{(+)\dagger}(\vec{x}) \big| A_{FW}^{(+)} \big| \varphi_{D}^{(+)}(\vec{x}) \right\rangle \\ \left\langle \psi_{FW}^{(-)\dagger}(\vec{x}) \big| A_{FW} \big| \psi_{FW}^{(-)}(\vec{x}) \right\rangle = \left\langle \chi_{D}^{(-)\dagger}(\vec{x}) \big| A_{FW}^{(-)} \big| \chi_{D}^{(-)}(\vec{x}) \right\rangle \end{array} ; \qquad (41)$$

In expressions (41) the four-component operator $A_{FW}$ is written in its matrix form

$$A_{FW} = \begin{pmatrix} A_{FW}^{(+)} & (A_{FW})_{12} \\ (A_{FW})_{21} & A_{FW}^{(-)} \end{pmatrix}.$$

The author gratefully acknowledges discussions with A.Ya.Silenko, who also draw attention to the Eriksen transformation [12] that allowed the author an easier proof of the sufficiency condition for transformation to the Foldy-Wouthhysen representation.